\documentclass[letterpaper, 10pt, conference]{IEEEtran}
\usepackage[top=0.75in, bottom=0.78in, left=0.75in, right=0.75in]{geometry}

\IEEEoverridecommandlockouts  

\usepackage{orcidlink}
\usepackage[normalem]{ulem} 
\usepackage{color}
\usepackage{tcolorbox}
\usepackage{listings}
\usepackage{xcolor} 
\usepackage{cite}
\usepackage{amsmath,amssymb,amsfonts}
\usepackage{algorithmic}
\usepackage{graphicx}
\usepackage{algorithm}
\usepackage{textcomp}
\usepackage{rotating}
\usepackage{multicol}
\usepackage{float}
\usepackage{pifont}
\usepackage{listings}
\usepackage{xcolor}
\usepackage{booktabs,tabularx}

\usepackage{stfloats}

\usepackage{todonotes}
\usepackage{pdflscape}
\usepackage{xcolor}
\def\BibTeX{{\rm B\kern-.05em{\sc i\kern-.025em b}\kern-.08em
    T\kern-.1667em\lower.7ex\hbox{E}\kern-.125emX}}

\usepackage{booktabs} 
\usepackage{fullpage} 
\usepackage{multirow}
\usepackage{graphicx} 
\usepackage[table]{xcolor} 

\lstdefinelanguage{json}{
  basicstyle=\ttfamily\footnotesize,
  numbers=left,
  numberstyle=\tiny,
  stepnumber=1,
  numbersep=8pt,
  showstringspaces=false,
  breaklines=true,
  frame=single,
  backgroundcolor=\color{gray!10},
  literate=
   *{0}{{{\color{blue}0}}}{1}
    {1}{{{\color{blue}1}}}{1}
    {2}{{{\color{blue}2}}}{1}
    {3}{{{\color{blue}3}}}{1}
    {4}{{{\color{blue}4}}}{1}
    {5}{{{\color{blue}5}}}{1}
    {6}{{{\color{blue}6}}}{1}
    {7}{{{\color{blue}7}}}{1}
    {8}{{{\color{blue}8}}}{1}
    {9}{{{\color{blue}9}}}{1}
    {:}{{{\color{red}{:}}}}{1}
    {,}{{{\color{red}{,}}}}{1}
    {"}{{{\color{green}{\texttt{"}}}}}{1}
}
\usepackage{soul}
\sethlcolor{lightgray!50}
\usepackage{multirow}
\usepackage{booktabs}
\usepackage[utf8]{inputenc}
\usepackage{tikz}
\usetikzlibrary{positioning, shapes, arrows.meta, fit}
\definecolor{mainblue}{RGB}{0,70,190}
\definecolor{lightbeige}{RGB}{255,250,240}

\usepackage{float}

\usepackage{hyperref}
\usepackage{soul} 

\definecolor{lightblue}{rgb}{0.93,0.96,1}
\definecolor{hellgrauHintergrundZwei}{rgb}{0.95, 0.95, 0.95}
\definecolor{dunkelblaugruen60heller}{rgb}{0.6, 0.78, 0.9}

\usepackage{amsmath,amssymb,amsfonts}
\usepackage{algorithmic}
\usepackage{graphicx}
\usepackage{textcomp}
\usepackage{xcolor}
\def\BibTeX{{\rm B\kern-.05em{\sc i\kern-.025em b}\kern-.08em
    T\kern-.1667em\lower.7ex\hbox{E}\kern-.125emX}}
\begin{document}

\title{
\vspace*{0.15in} 
PriMod4AI: Lifecycle-Aware Privacy Threat Modeling for AI Systems using LLM

}

\author{
    \IEEEauthorblockN{
        Gautam Savaliya\IEEEauthorrefmark{1}\orcidlink{0009-0008-8400-6368}, 
        Robert Aufschläger\orcidlink{0009-0004-0986-3504}, 
        Abhishek Subedi\orcidlink{0009-0004-2114-1360}, 
        Michael Heigl\orcidlink{0000-0001-7303-113X}, and 
        Martin Schramm\orcidlink{0000-0001-6206-2969}
    }
    \IEEEauthorblockA{
        Deggendorf Institute of Technology, Germany\\
        \IEEEauthorrefmark{1}gautam.savaliya@th-deg.de
    }
}

\maketitle
\begin{abstract}

Artificial intelligence systems introduce complex privacy risks throughout their lifecycle, especially when processing sensitive or high-dimensional data. Beyond the seven traditional privacy threat categories defined by the LINDDUN framework, AI systems are also exposed to model-centric privacy attacks such as membership inference and model inversion, which LINDDUN does not cover. To address both classical LINDDUN threats and  additional AI-driven privacy attacks, PriMod4AI introduces a hybrid privacy threat modeling approach that unifies two structured knowledge sources, a LINDDUN knowledge base representing the established taxonomy, and a model-centric privacy attack knowledge base capturing threats outside LINDDUN. These knowledge bases are embedded into a vector database for semantic retrieval and combined with system level metadata derived from Data Flow Diagram. PriMod4AI uses retrieval-augmented and Data Flow specific prompt generation to guide large language models (LLMs) in identifying, explaining, and categorizing privacy threats across lifecycle stages. The framework produces justified and taxonomy-grounded threat assessments that integrate both classical and AI-driven perspectives. Evaluation on two AI systems indicates that PriMod4AI provides broad coverage of classical privacy categories while additionally identifying model-centric privacy threats. The framework produces consistent, knowledge-grounded outputs across LLMs, as reflected in agreement scores in the observed range. 
\end{abstract}


\section{Introduction}
AI systems are increasingly deployed across domains such as healthcare, mobility, finance, and public services, where they process sensitive personal data including biometric identifiers, behavioral patterns, and contextual information. Such processing introduces substantial privacy risks, and any misuse or leakage may violate core regulatory principles such as lawfulness, data minimization, and purpose limitation defined in Article~5 of the GDPR. Traditional privacy threat modeling frameworks, such as LINDDUN (Linking, Identifying, Non-repudiation, Detecting, Data Disclosure, Unawareness, Non-compliance)~\cite{b1}, provide a well established taxonomy for identifying data-centric privacy risks in conventional software systems. However, they do not fully account for the dynamic, iterative, and model-driven nature of modern AI pipelines.

\begin{figure}[htbp]
\centerline{\includegraphics[width=6.8cm, height=4.8cm]{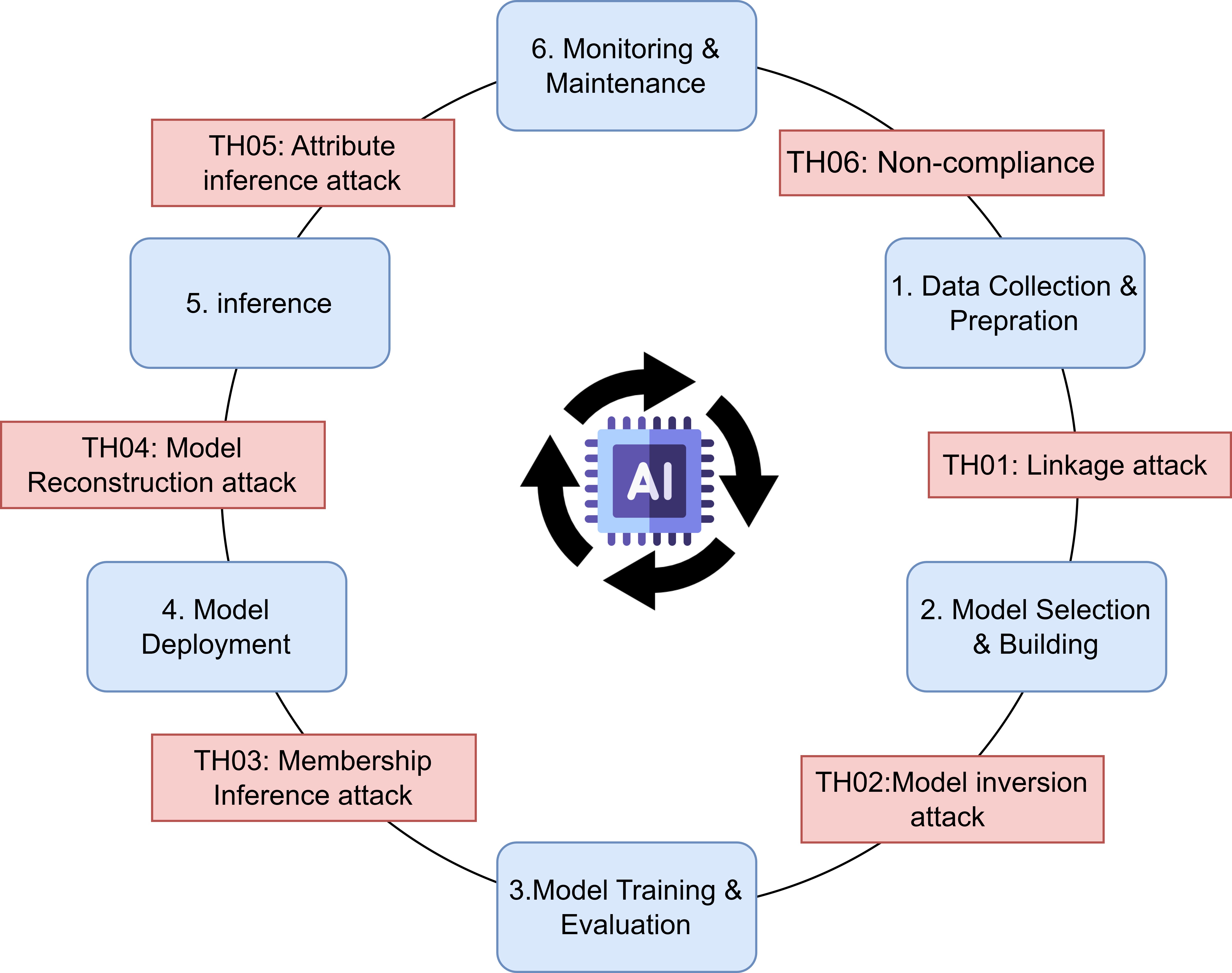}}
\caption{A six-phase AI development lifecycle encompassing data collection, model building, training, deployment, inference, and continuous monitoring. The diagram maps distinct privacy risks (shown in red boxes) to their corresponding stages within the lifecycle.}
\label{fig:ai_lifecycle}
\end{figure}

Beyond these, modern AI systems introduce a distinct class of model-centric privacy attacks rooted in the behavior of trained models. Such attacks exploit memorization, overfitting, or unintended information leakage from learned representations~\cite{b1.2, b1.3}. For example, a face-recognition model may inadvertently memorize training images, enabling membership inference~\cite{b1.4}, or reveal sensitive attributes through model inversion~\cite{b1.5}. Beyond these, AI models are also vulnerable to attribute inference attacks that predict hidden personal traits, training data extraction attacks~\cite{b1.2, b1.7} that recover specific records from generative and encoder–decoder architectures~\cite{b1.8}, shadow-model reconstruction techniques that approximate private datasets, and embedding-space leakage where vector representations disclose identifying or sensitive information~\cite{b1.9, b1.95}. These risks manifest across the entire AI lifecycle, from data collection and preprocessing to training, deployment, inference, and continuous monitoring~\cite{b1.1} as illustrated in Figure~\ref{fig:ai_lifecycle}.

Conventional privacy threat modeling techniques struggle to capture this dual landscape: LINDDUN-based approaches effectively identify classical privacy threats but overlook model-centric vulnerabilities, while automated extensions such as PILLAR\cite{b10.1} remain limited to the classical LINDDUN taxonomy and lack lifecycle awareness. This gap underscores the need for privacy threat-modeling methods that can jointly reason about both system-level and model-level privacy risks in AI systems.

The identified limitation underscores the need of lifecycle-aware methodology which are capable of addressing both traditional LINDDUN threats and emerging model-centric privacy attacks. PriMod4AI fulfills this need through a lifecycle-aware threat-identification pipeline that combines structured privacy knowledge bases with system metadata and employs LLM-driven retrieval-augmented generation to automate and explain privacy threat analysis.
Motivated by this, we investigate the following research questions:
\begin{itemize}
    \item \textbf{RQ1:} To what extent, PriMod4AI can extend traditional privacy threat-modeling approaches by jointly identifying data-centric LINDDUN threats and model-centric privacy attacks across the AI lifecycle?
    \item \textbf{RQ2:} How effectively can retrieval-augmented and knowledge-grounded LLM reasoning achieve reliable, consistent, and explainable privacy threat identification across multiple AI domains? 
\end{itemize}

\noindent Motivated by these research questions, our work offers the following contributions:

\begin{enumerate}
    \item \textbf{Dual structured privacy knowledge bases:} We develop two complementary knowledge sources: a LINDDUN knowledge base capturing the classical seven LINDDUN privacy threat categories, and a model-centric privacy attack knowledge base representing AI-driven threats such as membership inference, model inversion, and training data extraction.
    
    \item \textbf{Data Flow-Specific Retrieval-Augmented Prompting:} We embed both knowledge bases into a vector database for semantic retrieval and combine the retrieved knowledge with system-level metadata derived from DFD. This enables PriMod4AI to generate context-rich, DF-specific prompts that provide grounded and lifecycle-aware threat reasoning. 
    
    \item \textbf{Explainable LLM-based threat identification:} Leveraging open-source LLMs, PriMod4AI produces structured threat assessments that integrate LINDDUN categories, AI-lifecycle stages, and explicit knowledge-source attribution, improving explainability and interpretability. 
\end{enumerate}

\section{Related Works}

\subsection{Privacy Threat Modeling}
LINDDUN is a well known established privacy threat-modeling methodology for software systems, offering a structured taxonomy of seven privacy threat categories~\cite{b2}. Although several extensions, such as domain-specific refinements~\cite{b3}, improve its applicability, LINDDUN remains primarily design-time oriented and does not address privacy risks that emerge during model training, inference, or deployment. PLOT4AI provides an 86-threat catalog for AI technologies~\cite{b30}, yet its questionnaire-driven elicitation lacks lifecycle alignment and systematic mapping to system architecture. PILLAR automates LINDDUN analysis using LLMs~\cite{b10.1}, demonstrating the feasibility of LLM-assisted privacy threat modeling, but its reasoning is constrained to the classical taxonomy and does not incorporate model-centric threats or retrieval mechanisms. Together, these works establish the foundations for structured privacy analysis in software and data-processing systems. 

\subsection{AI-Specific Privacy Risks and Taxonomies}
AI systems introduce privacy threats that extend beyond traditional data-flow centric frameworks. Membership inference attacks can reveal whether an individual's data was used during training~\cite{b15}, while model inversion techniques reconstruct sensitive attributes from model outputs~\cite{b27}.
Recent work further shows that large generative models can memorize and leak training data or personally identifiable information~\cite{b1.2}, underscoring risks that arise not only during data collection but also during training and deployment.
Lifecycle analyses highlight the recurrence of privacy risks across evolving AI pipelines~\cite{b1.2,b37}.
Broader security oriented taxonomies such as ENISA’s AI Threat Landscape and NIST’s AI RMF~\cite{b19,b12} catalog assets and attack surfaces, but provide limited granularity for privacy-specific risks and do not link them to LINDDUN or lifecycle stages. These findings highlight that modern AI systems introduce privacy threats that needs to be addressed.

\subsection{Automated Threat Identification with LLMs}
Recent research applies LLMs for automating security and privacy analysis. ThreatGPT~\cite{b33} and ThreatModeling-LLM~\cite{b32} use generative prompting and fine-tuning to automate STRIDE/NIST-based threat elicitation, while Auspex~\cite{b21} and ThreatFinderAI~\cite{b22} incorporate expert knowledge or knowledge graphs for asset-centric reasoning. However, these systems primarily target cybersecurity threats and lack explicit privacy taxonomies or lifecycle grounding. PILLAR~\cite{b10.1} automates LINDDUN classification but relies solely on prompt instructions and cannot identify model-centric attacks such as inversion or membership inference. These approaches demonstrate the growing potential of LLMs for automating aspects of privacy and security assessment


\subsection{Retrieval-Augmented Generation in Threat Modeling}
Retrieval-Augmented Generation (RAG) has emerged as a powerful technique to enhance LLM-based analysis by grounding outputs in external knowledge sources, reducing hallucinations and improving factual accuracy~\cite{b3.1}. In threat modeling, recent frameworks leverage RAG to automate elicitation processes. For instance, ThreatLens~\cite{b3.2} integrates RAG with LLMs to generate threat models and test plans for hardware security verification, drawing from vulnerability databases. Similarly, a study shows that integrating retrieval mechanisms with generative models strengthens their ability to handle complex, information-dense queries. Such retrieval-augmented systems have been shown to provide more reliable and context-aware outputs, making them well-suited for domains where precision and up-to-date knowledge are essential~\cite{b3.3}. MoRSE~\cite{b3.4}, a cybersecurity chatbot, uses a mixture of RAG systems to provide comprehensive knowledge coverage across threat landscapes. These studies show that RAG can enhance domain reasoning in complex threat landscapes by grounding LLMs in structured external knowledge.
\par Although existing research provides valuable foundations across classical privacy taxonomies, AI-specific threat analyses, LLM-assisted modeling, and RAG-based reasoning, these efforts remain fragmented and insufficient for lifecycle-aware privacy-by-design. LINDDUN and its refinements focus on design-time, data-centric harms and do not capture threats emerging during model training or inference. AI-oriented taxonomies offer detailed accounts of model-centric attacks but lack integration with DFD semantics or privacy engineering workflows. Automated approaches such as PILLAR are the closest to our objectives, yet they remain limited to the classical LINDDUN space and cannot identify model-centric risks such as membership inference, inversion, or training-data leakage. Thus, they only partially align with the requirements of modern AI systems. Likewise, RAG-based threat modeling frameworks primarily address general cybersecurity vulnerabilities rather than privacy-specific harms, and do not incorporate structured knowledge bases or lifecycle mappings. These limitations collectively motivate PriMod4AI, which uniquely combines retrieval-augmented prompting, dual structured knowledge bases (LINDDUN and AI-specific threats), and DFD-level metadata to enable lifecycle-aware, explainable, and domain-consistent privacy threat identification across both classical and AI-specific threat spaces.

\section{Proposed Method}
\begin{figure*}[htbp]
\centerline{\includegraphics[width=\textwidth,height=7.5cm]{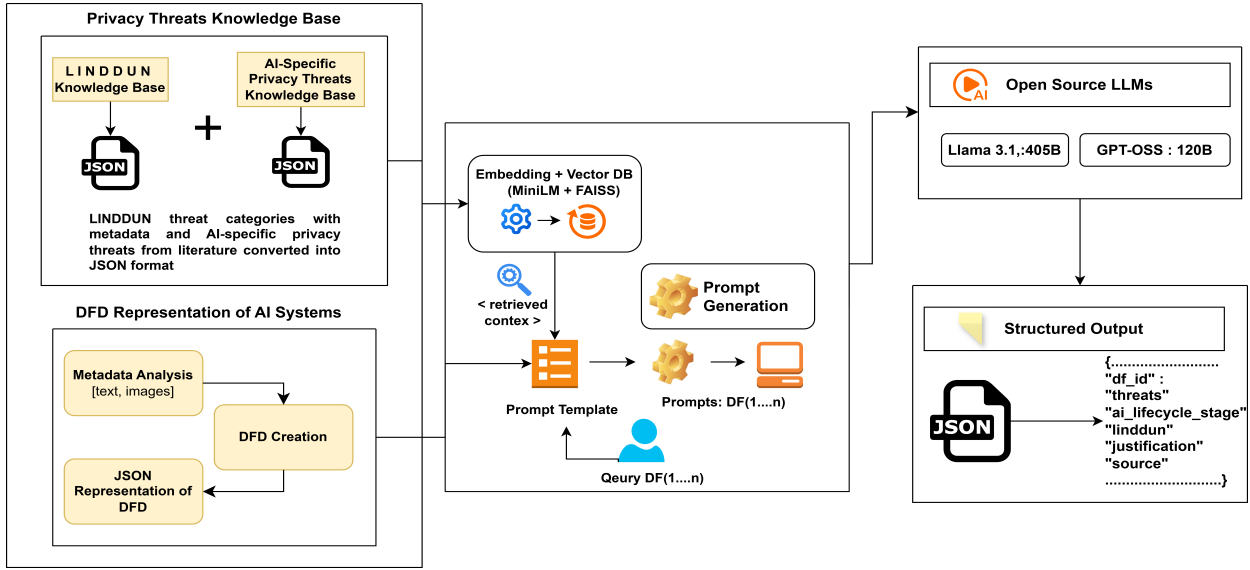}}
\caption{Architecture of the proposed PriMod4AI framework for automated privacy threat modeling in AI systems. The framework integrates a LINDDUN + AI-specific privacy threat knowledge base, DFD representation of AI systems, and open-source LLMs for prompt-based threat identification, producing structured JSON outputs.}
\label{fig:architecture_primod4ai}
\end{figure*}

This section outlines the methodological framework developed for automated, lifecycle-aware privacy threat modeling in AI systems. The overall architecture of the proposed framework is illustrated in Figure~\ref{fig:architecture_primod4ai}. By integrating domain knowledge and system-level representations, the framework enables contextual, LLM-driven threat reasoning. The methodology comprises five phases, beginning with Knowledge Base Construction, followed by DFD Representation of the AI System, Retrieval-Augmented Prompt Generation, and LLM Integration and Inference, and concluding with Structured Output Generation.

\subsection{Knowledge Base Construction}
The first phase of the proposed framework involves constructing two complementary knowledge bases to enable automated, context-aware privacy threat identification.

\textbf{LINDDUN Knowledge Base (LINDDUN\_KB):}
The official LINDDUN taxonomy, originally provided as a set of hierarchical PDF descriptions, was fully converted into a structured JSON-based knowledge base covering all seven LINDDUN privacy threat categories and their subnodes. Using NLP-assisted extraction followed by manual refinement, we preserved the complete hierarchy, including examples, criteria, impacts, and additional context for each threat node. Representing the taxonomy in JSON offers two key advantages: (i) it enables consistent and fine-grained retrieval of relevant threat information during prompt construction, and (ii) it allows LLMs to consume semantically structured content rather than raw PDF text, reducing ambiguity and improving grounding. 

\textbf {AI Model-Centric Privacy Attack Knowledge Base (AI\_Privacy\_KB):}
To extend coverage beyond traditional software privacy domains, an AI\_Privacy\_KB was developed through a structured, transparent literature review tailored to privacy risks unique to AI and ML. The review covered publications from 2016–2025 and examined peer-reviewed venues (\textit{IEEE Xplore}, \textit{ACM Digital Library}, \textit{SpringerLink}) alongside screened preprints from \textit{arXiv} to capture emerging threats not yet formally published. Searches used keyword combinations of \emph{AI privacy}, \emph{model inversion}, \emph{membership inference}, \emph{privacy attacks}, and \emph{threat modeling}, with inclusion criteria requiring that a source (i) described AI-lifecycle-specific privacy risks, (ii) articulated mitigation, governance, or regulatory implications (e.g., GDPR, AI Act), and (iii) provided adequate technical detail for structured threat encoding. After filtering and deduplication, about 30 unique peer-reviewed and regulatory sources remained. Each identified threat was reviewed, mapped to its closest LINDDUN dimension, and encoded in a JSON schema capturing its \texttt{name}, \texttt{description}, \texttt{attack vector}, \texttt{AI lifecycle stage}, and \texttt{source reference}, ensuring comprehensive and taxonomy-aligned coverage. The complete list of reviewed sources and corresponding threat mappings is provided in Appendix\ref{sec:AIprivacyKB}.

\subsection{DFD Representation of the AI System}
To support structured privacy reasoning within the PriMod4AI pipeline, the graphical DFD of the target AI system is first transformed into comprehensive textual and semantic representation. This process begins with metadata extraction, where system documentation, component descriptions, and signal specifications are analyzed to identify functional units, their interconnections, and the type and sensitivity of data exchanged between them. The analysis produces a detailed inventory of:  
(i) external entities,  
(ii) processes and processing functions,  
(iii) data stores,  
(iv) data flows, and  
(v) trust boundaries delineating privacy-relevant domains.
\begin{lstlisting}[language=json, numbers=none, caption={JSON representation of Data Flow DF1 (Camera → Sensor Fusion), illustrating how PriMod4AI encodes data-flow metadata including source, destination, data type, sensitivity, and lifecycle stage for downstream threat-analysis prompting and knowledge retrieval.}, label={lst:df1}]
"data_flows": [
  {
    "id": "DF1",
    "source": "E1",
    "destination": "P1",
    "data_type": "camera images/video",
    "sensitive_info": "visual scene data",
    "description": "Transfer of camera data to sensor fusion.",
    "lifecycle_stage": "Data Collection to Data Processing"
  }
]
\end{lstlisting}
Each data flow is then formalized into a JSON-based representation that captures its source, destination, data type, sensitivity classification, functional description, and associated AI lifecycle stage. An example snippet for Data Flow~DF1 (Camera~→~Sensor Fusion), taken from the AI system used in our experimentation and illustrated in Figure~\ref{fig:autonomous-dfd}, is provided in Listing~\ref{lst:df1}. These structured representations form the basis for generating DF-specific prompts and for retrieving relevant knowledge during later stages of the PriMod4AI pipeline.

\subsection{Retrieval-Augmented Prompt Generation}
\label{sec:prompt_generation}
This phase generates a structured prompt for each data flow by combining the JSON-encoded DFD metadata with relevant knowledge retrieved from the LINDDUN knowledge base and AI\_Privacy\_KB. A base prompt template is then assembled  with this information, producing DF–specific prompts that blend system context with taxonomy-grounded and model-centric threat knowledge.

\paragraph{Base Prompt Template}
The framework employs a structured base prompt template that defines the LLM’s analytic role, incorporates data-flow metadata, and specifies the required JSON output schema. An abbreviated version is shown in Listing~\ref{lst:base_prompt_template1}, the full template is included in Appendix\ref{sec:base_prompt_template}.

\begin{lstlisting}[language=json, numbers=none, caption={Abbreviated base prompt template guiding LLM-driven privacy threat identification. The template encodes analytic instructions, data-flow metadata, and output schema constraints.}, label={lst:base_prompt_template1}]
Analyze the following Data Flow {df_id} and identify privacy threats using retrieved knowledge from the LINDDUN_KB and AI_Privacy_KB.

### Data Flow:
{source} -> {destination}, {data_type}, {sensitive_info}
### Knowledge Context:
{context}
Return a JSON object with: name, justification, linddun_category,
ai_lifecycle_stage, and source.
\end{lstlisting}

Serving as the foundation of PriMod4AI's threat-identification process, this template standardizes how system metadata and retrieved knowledge are combined, ensuring consistent, explainable, and reproducible outputs across all data flows.

\paragraph{Per-DF Prompt Construction}
For each data flow ($DF_i$), a composite prompt ($P_i$) is instantiated from the base template by injecting DF-specific metadata including \texttt{id}, \texttt{source}, \texttt{destination}, \texttt{data\_type}, \texttt{sensitive\_info}, and \texttt{lifecycle\_stage} into predefined placeholders. This ensures that each prompt is precisely aligned with the semantics of its corresponding data flow while preserving a uniform structure across the pipeline.

Formally, given a base prompt $P_0$ and metadata set $m_i$, the composite prompt is defined as:
\begin{equation}
    P_i = f(P_0, m_i),
\end{equation}
where $f$ denotes the template-filling function.

After instantiation, each $P_i$ is enriched with retrieved knowledge via RAG before being provided to the LLM. This modular separation between the static template and DF-specific instantiation improves scalability by allowing new data flows to be analyzed without redesigning prompts while also enhancing reproducibility, since each generated prompt is deterministically constructed from explicit metadata and a fixed template.

\paragraph{Retrieval-Augmented Generation}
To provide factual grounding and reduce hallucination, PriMod4AI integrates a RAG pipeline that couples semantic retrieval with LLM-based reasoning. Both the LINDDUN\_KB and AI\_Privacy\_KB are transformed into structured JSON and embedded as dense vectors using the \texttt{MiniLM-L6-v2}~\cite{b2.1}  model from the Sentence Transformers family. The resulting embeddings are indexed using \texttt{FAISS}~\cite{b2.2}  for efficient nearest-neighbor retrieval. To maintain concept-level granularity, the knowledge bases are segmented via a recursive text-splitting strategy, ensuring that each node of the taxonomy remains independently retrievable.

For each instantiated prompt $P_i$, the textual description of the corresponding data flow denoted as $d_i$ and obtained directly from the DFD metadata, is encoded into a query vector $q_i$. The retriever then selects the top-$k$ ($k=7$) most semantically relevant knowledge fragments:
\begin{equation}
    S_i = \text{top-}k(R(q_i, K)),
\end{equation}
where $K$ denotes the embedded knowledge corpus and $R$ is the retrieval function.

The retrieved fragment set $S_i$ is inserted into the prompt’s \texttt{<context>} section, after which the remaining instruction block of the base template is appended. Formally, the augmented prompt is expressed as:
\begin{equation}
    \tilde{P}_i = \text{assemble}(P_i, S_i),
\end{equation}
where \texttt{assemble} denotes the template-aware integration of (i) DF-specific metadata, (ii) retrieved knowledge, and (iii) task instructions.

This approach grounds the LLM with authoritative, flow-specific context, preserves a fixed JSON output schema, and assigns each identified threat a source label from the annotated context to improve transparency and interpretability.

\subsection{Open-Source LLM Integration}
The next stage of the framework processes retrieval-enhanced prompts using open-source LLMs for automated privacy threat identification. To ensure transparency and reproducibility, only openly available models were employed via the \texttt{OllamaLLM} interface. Two variants were used: \texttt{GPT-OSS (120B)}\footnote{Ollama, “gpt-oss,” Ollama Library. [Online]. Available: https://ollama.com/library/gpt-oss. [Accessed: Dec. 3, 2025].} 
optimized for structured reasoning, and \texttt{LLaMA 3.1 (405B)}\footnote{AI at Meta, Jul. 23, 2024. [Online]. Available: https://ai.meta.com/blog/meta-llama-3-1/. [Accessed: Dec. 3, 2025].}.
offering enhanced contextual precision. Each augmented prompt $\tilde{P}_i$, containing system metadata and retrieved knowledge fragments, is supplied to the selected model for inference in schema-constrained JSON mode, ensuring a uniform output structure (\texttt{df\_id}, \texttt{identified\_threats}). Decoding parameters (e.g., temperature, top-p) are fixed across models to maintain comparability. This setup supports interchangeable inference and cross-model evaluation of reasoning consistency, coverage, and reproducibility.

\subsection{Structured Output Generation}
The final stage produces a structured output summarizing all identified privacy threats per DF.
\begin{lstlisting}[
    language=json,
    numbers=none,
    caption={The structured JSON output presented is for Data Flow 5 (DF5), which is derived from the AI-based system diagrammed in Figure \ref{fig:dfd_auth}. This result, generated by LLaMA 3.1 within the PRIMOD4AI framework, demonstrates the unified threat-encoding schema applied to an internal data flow of the analyzed AI system.},
    label={lst:structured_output_example},
    basicstyle=\small\ttfamily % Optional: Makes the font smaller and monospaced
]
{
"df_id": "DF5",
"identified_threats":[ {
    "name": "Unencrypted Data Transfer",
    "justification": "Sensitive biometric data may be exposed if transferred without encryption.",
    "linddun_category": "Disclosure of information",
    "ai_lifecycle_stage": "Inference/Storage",
    "source": "LINDDUN"
    },
    {
    "name": "Model Inversion Attack",
    "justification": "Stored embeddings could be exploited to reconstruct facial traits.",
    "linddun_category": "Disclosure of information",
    "ai_lifecycle_stage": "Inference",
    "source": "AI_PRIVACY_KB"
    }]}
\end{lstlisting}
For each DF, the LLM returns a standardized JSON record with threat names, LINDDUN categories, AI lifecycle stages, and source references. As shown in Listing~\ref{lst:structured_output_example}, this uniform format supports consistency, evaluation, and automated visualization, while enabling comparison and reproducibility across DFs.

\subsection{Algorithmic Pipeline for Threat Identification}

\begin{algorithm}[htbp]
\caption{PriMod4AI: Retrieval-Augmented LLM-Based Privacy Threat Identification}
\label{alg:rag_df}
\begin{algorithmic}[1]
\REQUIRE LINDDUN\_KB $\mathcal{K}_L$, AI\_Privacy\_KB $\mathcal{K}_A$, DFD JSON $\mathcal{D}$, base prompt template $\mathcal{P}$; embedding model $M_{\text{emb}}$; FAISS index $\mathcal{V}$ (initially empty); LLM $M_{\text{LLM}}$; retriever top-$k$ (here $k=7$).
\ENSURE Result list $\mathcal{R}$ of zero or more JSON objects of the form \texttt{\{"df\_id": ..., "identified\_threats": [...]\}}.

\STATE $\mathcal{R} \leftarrow [\ ]$ \COMMENT{Initialize result list}

\STATE \textbf{Build knowledge index:}
\STATE Split $\mathcal{K}_L$ and $\mathcal{K}_A$ into concept-level text chunks.
\FOR{each chunk $c$ in $\mathcal{K}_L \cup \mathcal{K}_A$}
    \STATE $e \leftarrow M_{\text{emb}}(c)$
    \STATE Insert $e$ into FAISS index $\mathcal{V}$
\ENDFOR

\STATE \textbf{Load inputs:}
\STATE Parse $\mathcal{D}$ into the set of data flows $\{DF_j\}$ and load the base prompt template $\mathcal{P}$.

\FOR{each data flow $DF_j$ in $\{DF_j\}$}
    \STATE Extract DF metadata $m_j$ (source, destination, data\_type, sensitive\_info, lifecycle\_stage) and textual description $d_j$.
    \STATE Construct query vector $Q_j \leftarrow M_{\text{emb}}(d_j)$.
    \STATE Retrieve relevant fragments: $S_j \leftarrow \text{top-}k(R(Q_j,\mathcal{V}))$.
    \STATE Form composite prompt $X_j \leftarrow \text{assemble}(\mathcal{P}, m_j, S_j)$.
    \STATE Generate JSON output $Y_j \leftarrow M_{\text{LLM}}(X_j)$.
    \STATE Validate $Y_j$ against the expected schema; optionally apply repair or re-prompting if invalid.
    \STATE Append $Y_j$ to $\mathcal{R}$.
\ENDFOR

\STATE \textbf{Store results:} Write $\mathcal{R}$ to \texttt{outputs}.

\end{algorithmic}
\end{algorithm}

\section{Experimental Setup}

This section describes the implementation setup and experimental procedure used to validate PriMod4AI. Two cross-domain AI-based systems, a Face Authentication System and an Autonomous Driving System were selected due to their distinct privacy sensitivities and heterogeneous data modalities.  The implementation outlines how PriMod4AI was executed end-to-end, detailing the experimental environment and computational resources used during inference. Table~\ref{tab:exp_setup} summarizes the complete environment configuration employed across all experiments.

\begin{table}[htbp]
\caption{Hardware and software configuration used for implementing and executing PriMod4AI.}
\label{tab:exp_setup}
\centering
\footnotesize
\renewcommand{\arraystretch}{1.15}
\begin{tabularx}{\columnwidth}{@{}l X@{}}
\toprule
\textbf{Component} & \textbf{Specification / Description} \\
\midrule
Host System & Windows~11; Virtual Machine \\
CPU & AMD Ryzen 5 5625U (VM configuration); 16\,GB RAM \\
GPU Hardware & NVIDIA A100 80\,GB PCIe (MIG profile: 3g.40\,GB) \\
Environment & Python~3.11; \texttt{OllamaLLM} inference interface \\
Model~1 & LLaMA~3.1 (405B), open-source dense decoder model \\
Model~2 & GPT-OSS (120B), open-source LLM for structured reasoning \\
Decoding Parameters & Temperature = 0.7; top-p = 0.9; max tokens = 1024 \\
\bottomrule
\end{tabularx}
\end{table}

\subsection{Use Cases}

\subsubsection{AI-based Face Authentication System}
\begin{figure}[htbp]
    \centering
    \includegraphics[width=7cm, height=5cm]{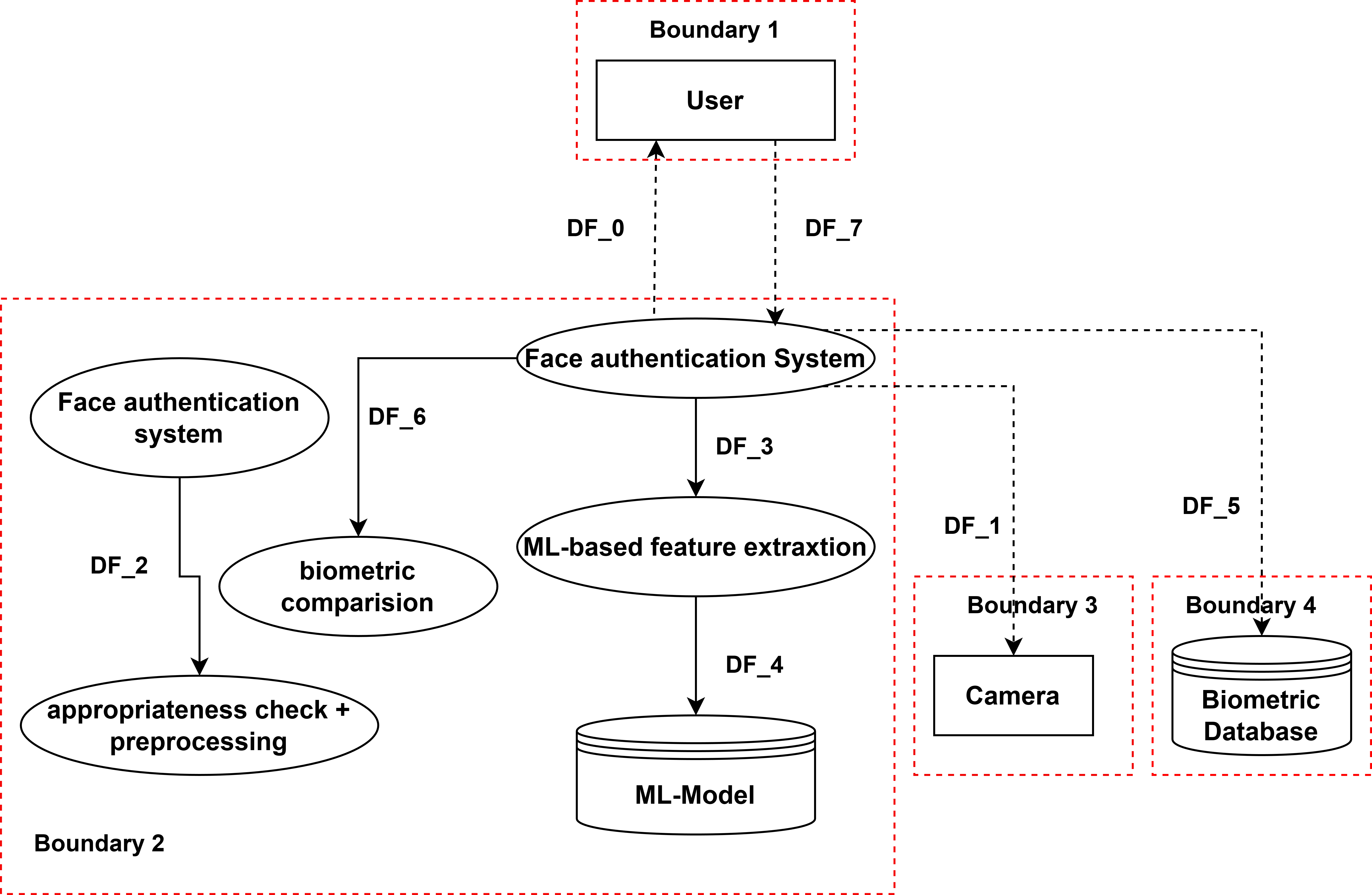}
    \caption{DFD of the AI-based Face Authentication System, adapted from the open-source PILLAR repository}
    \label{fig:dfd_auth}
\end{figure}

The first system represents a typical biometric verification pipeline that processes facial images, embeddings, and identity records. The dataset and reference architecture were adapted from the open-source PILLAR repository\footnote{PILLAR: LINDDUN Privacy Threat Modeling Using LLMs. [Online]. Available: https://github.com/stfbk/PILLAR (Accessed: Nov. 5, 2025).} ensuring compatibility with standard LINDDUN threat definitions. The DFD of the system is shown in Figure~\ref{fig:dfd_auth}.

The architecture consists of a camera (data source), preprocessing and feature extraction modules, a biometric comparison component, and storage units including databases and ML models. Each data flow (DF1–DF7) was analyzed to capture its source, destination, data type, sensitivity, and corresponding AI lifecycle stage.
\subsubsection{Autonomous Driving System}

\begin{figure}[htbp]
\centering
\includegraphics[width=8cm, height=5cm]{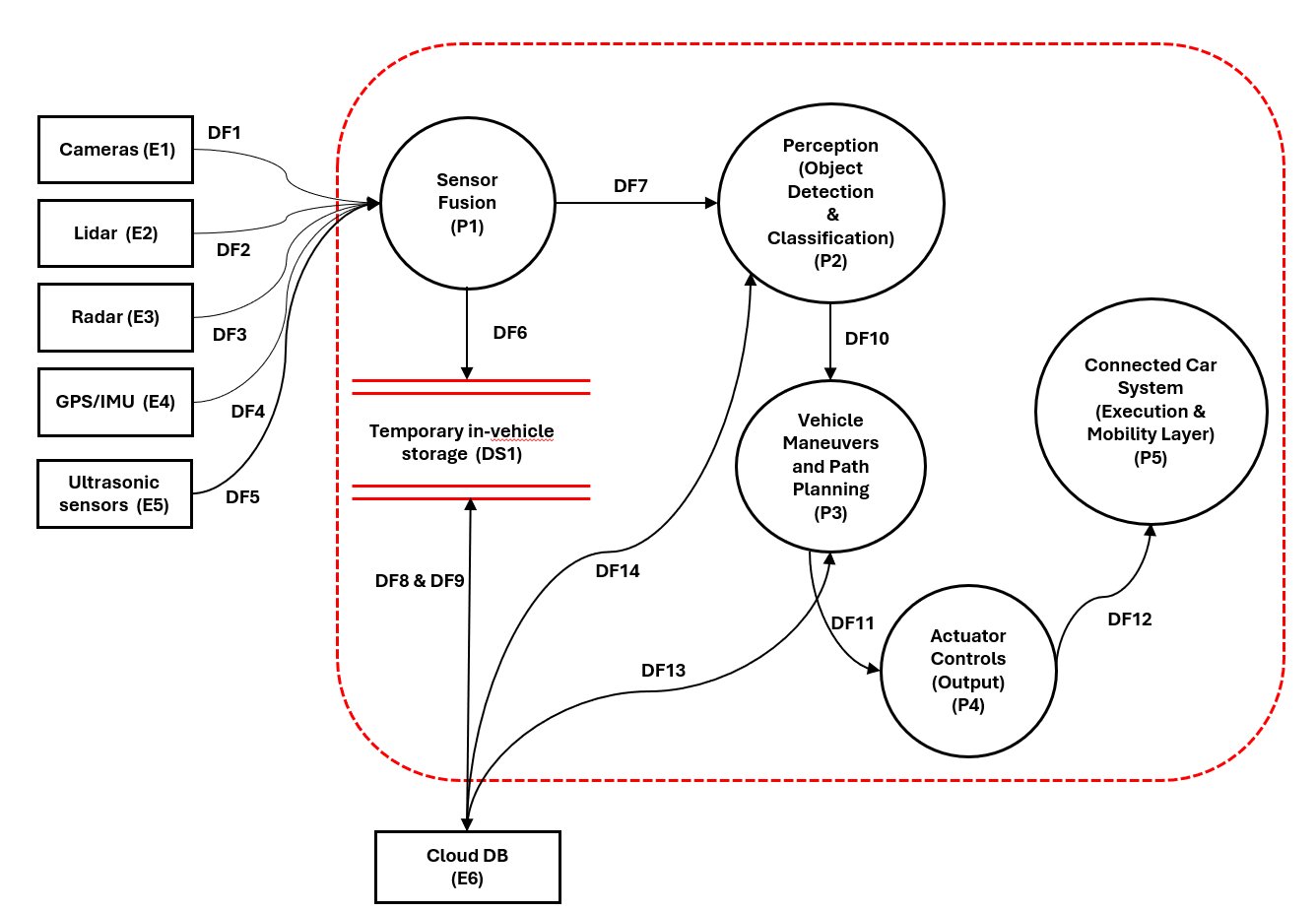}
\caption{DFD of the autonomous driving system showing key processes, data stores, and data flows (DF1–DF14).}
\label{fig:autonomous-dfd}
\end{figure}

The second system represents a generalized autonomous driving architecture derived from the government-funded JUST BETTER DATA (JBData) research project\footnote{just better DATA: Effiziente und hochgenaue Datenerzeugung für KI-Anwendungen im Bereich autonomes Fahren. [Online]. Available: https://www.justbetterdata.de/konzept/ (Accessed: Nov. 5, 2025).}. Due to the scale and complexity of the original system, and to ensure confidentiality, the architecture has been abstracted and generalized for experimental use. While specific implementation details are omitted, the core structural characteristics, sensor modalities, and representative data flows are preserved to maintain realism and suitability for privacy analysis.As shown in Figure~\ref{fig:autonomous-dfd}, the system integrates multi-sensor perception, fusion modules, planning and control components, and cloud-assisted synchronization processes. The architecture comprises fourteen data flows (DF1–DF14) spanning multiple AI lifecycle stages, providing a realistic and comprehensive testbed for evaluating lifecycle-aware privacy threat identification methods.

\section{Evaluation}
The evaluation assesses the performance and reliability of PriMod4AI using a two-layer structure, summarized in Table~\ref{tab:eval_layers}. The layered design reflects the dual nature of PriMod4AI’s outputs, which include both classical LINDDUN threats and AI-driven, model-centric privacy attacks.
\begin{table}[htbp]
\caption{Overview of Evaluation Layers and Associated Metrics.}
\label{tab:eval_layers}
\centering
\footnotesize
\renewcommand{\arraystretch}{1.1}
\begin{tabular}{p{3cm} p{4cm}}
\toprule
\textbf{Layer} & \textbf{Metrics Used} \\
\midrule
\textbf{Layer A: Classical LINDDUN Space} & 
\begin{enumerate}
\item Category Coverage 
\item PILLAR-Recall 
\item Jaccard Similarity
\end{enumerate} \\[2pt]

\textbf{Layer B: Cross-Model Agreement (LINDDUN + AI)} & 
\begin{enumerate}
\item Cohen’s~$\kappa$
    \begin{itemize}
        \item $\boldsymbol{P_o}$
        \item PABAK
    \end{itemize}
\item Robustness Coefficient
\end{enumerate} \\
\bottomrule
\end{tabular}
\end{table}

\subsection{Layer~A: Classical LINDDUN Threat Identification (Comparison with PILLAR)}
In this evaluation layer, we assess PriMod4AI, implemented using both the GPT-OSS and LLaMA 3.1 models within the classical LINDDUN privacy-threat space for both examined systems. Traditional LINDDUN modeling defines seven canonical threat categories (Linkability, Identifiability, Non-repudiation, Detectability, Disclosure of Information, Unawareness, and Non-compliance). To understand how PriMod4AI aligns with this taxonomy, we compare its category-level outputs against those produced by PILLAR, an LLM-based tool that automates the original LINDDUN workflow. PILLAR is not treated as a ground-truth baseline, but rather as a point of comparison for classical LINDDUN reasoning. The tool systematically generates analysis across all seven LINDDUN categories for each data flow for LINDDUN Pro method. 
\paragraph{Experimental Setup for PILLAR}
We executed PILLAR using its official Streamlit web application\footnote{https://pillar-ptm.streamlit.app/, (Accessed: Nov. 5, 2025) } with the GPT-4 Turbo model (OpenAI API), temperature $0.7$. The application requires structured input describing the system, data types, data transformations, actors, and data governance properties. For questions relating to data retention and deletion, we used a standardized organizational data policy. PILLAR then generated threats for all seven LINDDUN categories for each data flow (DF), after which a human analyst performed the relevance filtering step described above. PriMod4AI (LLaMA~3.1) and PriMod4AI (GPT-OSS) were evaluated against the filtered PILLAR results using the classical LINDDUN framework and the following metrics: 
\begin{itemize}
\item \textbf{Category Coverage (\%)~\cite{b2.5}:}

Measures how many of the seven LINDDUN categories are detected by a threat-identification model $m$ (PILLAR or PriMod4AI) for a specific data flow $d$ in the system’s DFD.

Let $\mathcal{L}=\{\ell_1,\ldots,\ell_7\}$ denote the LINDDUN category set. The coverage is defined as:
\begin{equation}
\mathrm{Cov}_m^{(d)} =
\frac{\sum_{\ell \in \mathcal{L}}
\mathbf{1}_{\text{threat}}(\ell, m, d)}
{|\mathcal{L}|}.
\end{equation}
Here, $\mathbf{1}_{\text{threat}}(\ell, m, d)$ equals $1$ if model $m$ identifies at least one threat in category $\ell$ for data flow $d$, and $0$ otherwise.
\item \textbf{PILLAR-Recall (Category-Based)\cite{b2.7, b2.8}:}  
Assesses backward compatibility by computing the fraction of (data-flow, 
category) pairs identified by PILLAR that are also identified by PriMod4AI.  
A recall of $1.0$ indicates perfect reproduction of PILLAR’s LINDDUN coverage.
\item \textbf{Jaccard Similarity (Per DF)\cite{b36}:}

Measures set-similarity between PriMod4AI and PILLAR on a per-data-flow basis:
\begin{equation}
J_m(d) =
\frac{|C_m(d) \cap C_{\mathrm{PILLAR}}(d)|}
{|C_m(d) \cup C_{\mathrm{PILLAR}}(d)|}.
\end{equation}
We report both per-DF scores and the average across all DFs. 
\end{itemize}

\subsection{Layer~B: Overall Privacy Threat Space and Cross-Model Agreement Analysis}
PriMod4AI identifies both classical LINDDUN threats and AI-driven, model-centric 
privacy attacks, forming an extended privacy threat space for which no 
benchmarks or expert-annotated ground truth exist. As direct accuracy-based 
evaluation is therefore infeasible. Layer~B assesses the reliability and factual consistency of PriMod4AI’s outputs by measuring the agreement between two LLMs (GPT-OSS and LLaMA3.1) across both evaluation systems and across all identified threat types. To ensure consistent agreement computation across models, threat names extracted from the structured JSON outputs are normalized using a preprocessing pipeline comprising: (i) lowercasing and lemmatization, (ii) stopword and punctuation removal, and (iii) token-set normalization. After preprocessing, semantically similar labels are merged using token-set Jaccard similarity, clustering two threat names when their similarity exceeds $\tau=0.20$, a commonly used threshold balancing precision and recall for near-duplicate textual concepts.


Three complementary statistical measures are then applied:

\begin{enumerate}

\item \textbf{Observed Agreement ($P_o$) \cite{b3.6}:}  
Represents the raw proportion of semantic threat clusters on which both models 
(GPT-OSS and LLaMA3.1) agree, either by jointly predicting the presence or the 
absence of a threat.  
It serves as the foundational quantity from which the other agreement metrics 
are derived.

\item \textbf{Cohen’s $\kappa$} \cite{b3.6}:  
A chance-corrected measure of agreement that quantifies how consistently the 
two models identify the same semantic threat clusters beyond what would be 
expected by random coincidence.  
For observed agreement $P_o$ and expected agreement $P_e$, Cohen's $\kappa$ is 
defined as:
\begin{equation}
\kappa = \frac{P_o - P_e}{1 - P_e}.
\end{equation}
Values above $0.75$ are typically interpreted as indicating substantial to 
near-perfect agreement.

\item \textbf{PABAK (Prevalence-Adjusted Bias-Adjusted Kappa)} 
\cite{b3.7}:  
A modified form of Cohen’s $\kappa$ designed for imbalanced binary data, where 
most threat clusters are absent.  
To correct the instability of $\kappa$ under low-prevalence conditions, PABAK 
adjusts the estimate using only observed agreement:
\begin{equation}
\text{PABAK} = 2P_o - 1.
\end{equation}
This yields a more robust agreement measure in sparse threat-identification 
settings.

\end{enumerate}

\section{Results and Discussion}
\subsection{Layer A: Classical LINDDUN Threat Space}
Layer~A evaluates PriMod4AI within the classical LINDDUN threat space by comparing its outputs to the PILLAR output as shown in Table~\ref{tab:results_all_layers}. Across both AI systems, PriMod4AI achieves moderate and acceptable agreement with PILLAR, as reflected by its PILLAR-Recall scores and Jaccard overlaps. The GPT-OSS variant consistently shows the closest alignment with PILLAR, achieving higher recall and broader category coverage, whereas the LLaMA3.1 variant provides a more compact and selective set of categories, resulting in slightly lower coverage but still maintaining reasonable overlap. The results also suggest that both PriMod4AI variants remain stable across heterogeneous system architectures, including the more complex multi-flow Autonomous Driving system. Overall, Layer~A shows that PriMod4AI remains compatible with classical LINDDUN reasoning while enabling structured, lifecycle-aware threat assessment. Appendix~\ref{sec:category_coverage} details per-flow LINDDUN analyses, reporting category coverage and identified threats for PILLAR, PriMod4AI (GPT-OSS), and PriMod4AI (LLaMA~3.1).

\subsection{Layer B: Combined Threat Space and Model-Centric Analysis}
Layer~B evaluates the consistency of PriMod4AI across different LLM variants by assessing cross-model agreement within the combined (LINDDUN + AI-specific) threat space. Since no ground-truth dataset exists for this domain, reliability is examined through inter-model metrics reported in Table~\ref{tab:results_all_layers}. Across both systems, the agreement scores between PriMod4AI (GPT-OSS) and PriMod4AI (LLaMA3.1) fall within a moderate to substantial range, indicating that the two models produce broadly comparable threat sets despite differences in wording or granularity. The observed agreement ($P_o$) and PABAK values suggest that both variants follow similar decision patterns rather than diverging arbitrarily, while Cohen’s~$\kappa$ show that the overlap between their outputs is meaningful but not identical. To summarize stability across system architectures, the normalized robustness coefficient $R$ indicates that PriMod4AI achieves moderate and consistent cross-model reliability, with $R = 0.7018$ for the face authentication system and $R = 0.6117$ for the autonomous driving system, reflecting stable but not exceptionally high robustness.Model-centric privacy risks are detailed in Appendix~\ref{sec:model_centric}, including their threat families and canonical categorization.


\begin{table*}[htbp]
\centering
\scriptsize
\setlength{\tabcolsep}{4.2pt}
\renewcommand{\arraystretch}{1.15}
\caption{Summary of evaluation results across the two analytical layers: Layer~A (Classical LINDDUN Threat Identification, compared with PILLAR) and Layer~B (Overall Privacy Threat Space and Cross-Model Agreement Analysis). Each section includes an interpretation column highlighting the primary observations.}
\label{tab:results_all_layers}
\begin{tabular}{l l c c c p{4.7cm}}
\toprule
\multicolumn{6}{l}{\textbf{Layer~A: Classical LINDDUN Threat Identification (Comparison with PILLAR)}}\\
\addlinespace[2pt]
\textbf{System} & \textbf{Model} & \textbf{PILLAR-Recall} & \textbf{Avg.\ Coverage} & \textbf{Avg.\ Jaccard} & \textbf{Interpretation} \\
\midrule

\multirow{3}{*}{Face Authentication} 
 & PILLAR & -- & 71.4\% & -- & Reference tool providing rule-based LINDDUN outputs used for comparison after analyst filtering.\\
 & PriMod4AI (GPT-OSS) & 85.0\% & 82.1\% & 0.642 & High alignment with PILLAR and broad category coverage, indicating strong classical LINDDUN consistency.\\
 & PriMod4AI (LLaMA)   & 77.5\% & 73.2\% & 0.617 & Moderate overlap with PILLAR with more selective category usage, reflecting a more compact reasoning style.\\

\midrule 

\multirow{3}{*}{Autonomous Driving}  
 & PILLAR & -- & 82.6\% & -- & Produces full-category threat suggestions requiring analyst filtering; used as comparative reference.\\
 & PriMod4AI (GPT-OSS) & 85.2\% & 81.1\% & 0.726 & Strong agreement with PILLAR and stable category coverage across complex multi-flow system.\\
 & PriMod4AI (LLaMA)   & 80.2\% & 71.3\% & 0.686 & Good overlap with PILLAR while exhibiting greater divergence in category selection across flows.\\

\midrule
\multicolumn{6}{l}{\textbf{Layer~B: Overall Privacy Threat Space and Cross-Model Agreement Analysis}}\\
\addlinespace[2pt]
\textbf{System} & \textbf{Model Pair} & \textbf{$\boldsymbol{\kappa}$} & \textbf{$\boldsymbol{P_o}$} & \textbf{PABAK} & \textbf{Interpretation} \\
\midrule
Face Authentication & GPT-OSS $\leftrightarrow$ LLaMA & 0.7455 & 0.7818 & 0.5782 & Moderate agreement levels, indicating generally consistent and largely non-hallucinatory reasoning..\\
Autonomous Driving  & GPT-OSS $\leftrightarrow$ LLaMA & 0.69 & 0.715 & 0.43 & Moderate cross-model consistency across diverse system architectures.\\
\bottomrule
\end{tabular}
\end{table*}

\section{\uppercase{Conclusion and Future Work}}

This work introduced PriMod4AI, a lifecycle-aware privacy threat-modeling framework that unifies structured privacy knowledge with retrieval-augmented LLM reasoning for automated, explainable privacy risk analysis in AI systems.

\textbf{RQ1} examined whether PriMod4AI can extend traditional privacy threat modeling by identifying both LINDDUN threats and AI-specific, model-centric risks. The results indicate that PriMod4AI maintains compatibility with classical LINDDUN outputs while additionally detecting a range of AI-driven threats not captured by PILLAR. This suggests that the framework can broaden the assessed privacy threat space in a structured manner.

\textbf{RQ2} investigated whether retrieval-augmented and knowledge-grounded prompting supports consistent and reliable threat identification. The cross-model agreement metrics reported in our evaluation show moderate to substantial alignment between the GPT-OSS and LLaMA3.1 variants, indicating that PriMod4AI produces generally stable and repeatable threat assessments across heterogeneous system architectures. 

These findings demonstrate that structured knowledge integration and RAG-based prompting provide a practical way to enhance coverage and reproducibility in privacy threat modeling for AI systems.PriMod4AI thus demonstrates that structured knowledge integration and RAG-based prompting can significantly enhance the rigor, coverage, and reproducibility of privacy threat modeling for the AI systems. Future work will focus on several practical extensions. First, the current evaluation is limited by the absence of expert-validated ground truth for AI-driven, model centric privacy threats so incorporating expert feedback loops could strengthen external validity. Second, the static knowledge bases may be expanded into lightweight, updateable knowledge repository to better track emerging privacy risks and regulatory changes. Third, integrating simple mitigation guidance for the identified threats would improve the framework’s usability in real development settings. Additionally, extending the pipeline to support incremental updates or domain-specific modules may further enhance adaptability across diverse AI architectures.

\section*{Acknowledgement}

The research leading to these results is funded by the German Federal Ministry for Economic Affairs and Energy within the project ``just better DATA - Effiziente und hochgenaue Datenerzeugung für KI-Anwendungen im Bereich autonomes Fahren''. The authors would like to thank the consortium for the successful cooperation.



\section*{APPENDIX}
\appendices

\subsection{Base Prompt Template}\label{sec:base_prompt_template}
\begin{lstlisting}[language=json, numbers=none, caption={Base prompt template used for automated privacy threat identification. The prompt integrates LINDDUN knowledge, AI-specific privacy risks, and structured system metadata to elicit grounded and machine-readable reasoning.}, label={lst:base_prompt_template}]
You are an expert privacy threat analyst specializing in AI systems. Your task is to analyze the provided Data Flow (DF) and identify all relevant privacy threats.
### Data Flow Context
- ID: {df_id}
- Source: {source}
- Destination: {destination}
- Data Type: {data_type}
- Sensitive: {sensitive}
- Sensitive Info: {sensitive_info}
- AI Lifecycle Stage: {lifecycle_stage}
### Knowledge Base (retrieved context)
{context}
### Threat Identification Rules (IMPORTANT)
1. Use BOTH:
   - your own expertise in privacy, LINDDUN, and AI-specific threats, AND
   - the retrieved context text above.
   The context is guidance, not a mandatory constraint.
2. Identify only those threats that are **logically possible** for THIS DF,
   based strictly on:
   - source and destination,
   - data_type,
   - sensitivity,
   - sensitive_info,
   - lifecycle_stage.
   Do NOT invent threats or include threats that contradict the DF.
3. LINDDUN threats:
   - Include a LINDDUN threat ONLY if it genuinely applies.
   - If `sensitive` = true, at least one LINDDUN threat is expected unless the DF truly poses no risk. 
4. AI-specific threats:
   Include an AI-specific threat **only if this DF involves. Do NOT include an AI-specific threat if none logically apply.
5. Avoid repeating identical threats across DFs unless the same threat clearly applies
   to the same scenario.
6. For each identified threat, include:
   - "name": a clear and specific threat name.
   - "justification": explain why this threat applies, referencing DF-specific fields
     (source, destination, data_type, sensitive_info, lifecycle_stage).
   - "linddun_category":  LINDDUN category.
   - "ai_lifecycle_stage": the lifecycle stage where the threat occurs
   - "source": "LINDDUN" for traditional privacy threats,
               "AI_PRIVACY_KB" for AI-specific threats.
\end{lstlisting}
The PriMod4AI framework relies on a structured and retrieval-augmented prompt design that standardizes how LLMs perform privacy threat analysis, as reflected in the complete base prompt template shown in Listing~\ref{lst:base_prompt_template}, ensuring uniform incorporation of retrieved context, lifecycle stages, and DFD attributes, thereby enabling systematic and well-grounded threat identification across AI systems. 

The prompt template is intentionally modular and role-based, enabling the LLM to combine system-level metadata, lifecycle context, and domain knowledge from both the LINDDUN taxonomy and the AI-specific privacy threat knowledge base. The template enforces a controlled reasoning process by (i) explicitly describing the data flow under analysis, (ii) injecting only the relevant subset of the knowledge bases retrieved for that flow,
The template also operationalizes dual-mode reasoning. First, it directs the LLM to apply the seven LINDDUN categories (Linkability, Identifiability, Non-repudiation, Detectability, Disclosure of Information, Unawareness, and Non-compliance) to the given data flow. Second, it supplements this with AI-specific, model-centric threat reasoning (e.g., membership inference, model inversion, model extraction, training-data leakage) sourced from the AI Privacy KB. When an AI threat does not directly map onto a LINDDUN category, the template requires the model to provide a justified mapping, thereby aligning emerging AI privacy risks with classical taxonomic structures. 

By integrating structured retrieval, explicit lifecycle grounding, and strict output constraints, this template enables reproducible, explainable, and cross-model-consistent privacy threat identification. It prevents unconstrained generative behavior, reduces hallucinations, and ensures that every reported threat is explicitly tied to a knowledge source and a lifecycle stage.
It prevents unconstrained generative behavior, reduces hallucinations, and ensures that every reported threat is explicitly tied to a knowledge source and a lifecycle stage.

\subsection{ Layer A: Classical LINDDUN Threat Space (LINDDUN Category Coverage Across Data Flows)}\label{sec:category_coverage}
To evaluate the breadth of classical privacy threat identification, Tables~\ref{tab:layerA_faceauth_primod_pillar_updated} and \ref{tab:autonomous_system_threats_updated} summarize, for each data flow in the studied systems, the set of LINDDUN categories detected by the evaluated methods. Rather than listing each category in a dedicated column, the tables present the activated category set together with its overall count. This compact representation enables a clear, flow-by-flow comparison among the three approaches: PILLAR, as an automated LINDDUN analysis tool, and the two PriMod4AI variants powered by GPT-OSS and LLaMA-3.1.

The results illustrate characteristic patterns in how each method engages with the LINDDUN taxonomy. The PriMod4AI models frequently identify a broad range of categories for individual data flows, reflecting their capacity to generate diverse, context-sensitive threat hypotheses. PILLAR, in turn, produces outputs that reflect its structured analytical process, providing consistent detection aligned with its underlying LINDDUN mapping logic. These differences do not indicate superiority of one approach over another; rather, they highlight the methodological distinctions between a deterministic LINDDUN engine and LLM-driven reasoning processes that are capable of exploring a wider interpretation space.

Taken together, the tables provide a unified view of category-level activation across the evaluated systems and methods. This presentation supports an assessment of how comprehensively each approach engages with the classical LINDDUN framework, without presupposing which style of analysis is preferable.

\begin{table*}[!t]
\centering
\caption{Layer-A LINDDUN threats identified for each data flow (DF0--DF7) in the face authentication system, comparing PriMod4AI (GPT-OSS and LLaMA variants) with PILLAR. 
Each cell lists the threats followed by the number of LINDDUN categories identified (in brackets).
Threat abbreviations: 
L = Linkability, I = Identifiability, DI = Disclosure of Information, DT = Detectability, 
U = Unawareness, Nr/NR = Non-repudiation, Nc/NC = Non-compliance.}
\label{tab:layerA_faceauth_primod_pillar_updated}
\begin{tabular}{l p{4.3cm} p{4.3cm} p{4.3cm}}
\hline
\textbf{Data Flow} & \textbf{PriMod4AI (GPT-OSS)} & \textbf{PriModAI(Llama3.1)} & \textbf{PILLAR} \\
\hline

DF0 & L, I, D, DD, U, Nc \textbf{[6]} 
    & L, I, D, DD, U, Nc \textbf{[6]} 
    & L, I, DD \textbf{[3]} \\

DF1 & L, I, D, DD, U, Nc \textbf{[6]} 
    & L, I, D, DD, Nr, U, Nc \textbf{[7]} 
    & L, I, D, DD \textbf{[4]} \\

DF2 & L, I, U, Nc \textbf{[4]} 
    & L, I, DD, U, Nc \textbf{[5]} 
    & L, I, D, DD, Nr, Nc \textbf{[6]} \\

DF3 & L, I, D, Nr, DD, U, Nc \textbf{[7]} 
    & L, I, D, DD, Nr, U, Nc \textbf{[7]} 
    & L, I, D, DD, Nr, Nc \textbf{[6]} \\

DF4 & L, I, Nr, U, Nc \textbf{[5]} 
    & L, I, D, DD, Nr, U, Nc \textbf{[7]} 
    & L, I, D, DD, Nr, Nc \textbf{[6]} \\

DF5 & L, I, DD, Nr, U, Nc \textbf{[6]} 
    & L, I, DD, Nr, U, Nc \textbf{[6]} 
    & L, I, D, DD, Nr, Nc \textbf{[6]} \\

DF6 & L, I, DD, Nc \textbf{[4]} 
    & L, I, DD, U, Nc \textbf{[5]} 
    & L, I, D, DD, Nc \textbf{[5]} \\

DF7 & D, DD, Nr \textbf{[3]} 
    & D, Nr, U \textbf{[3]} 
    & L, I, D, Nr \textbf{[4]} \\

\hline
\end{tabular}
\end{table*}

\begin{table*}[htbp]
\centering
\caption{LINDDUN threats identified for each data flow (DF1--DF14) in the Autonomous Driving System, comparing PriMod4AI (GPT-OSS and LLaMA variants) with PILLAR. 
Each cell lists the threats followed by the number of LINDDUN categories identified (in brackets).
Threat abbreviations: 
L = Linkability, I = Identifiability, DI = Disclosure of Information, DT = Detectability, 
U = Unawareness, NR = Non-repudiation, NC = Non-compliance.}
\label{tab:autonomous_system_threats_updated}
\begin{tabular}{l p{4.5cm} p{4.5cm} p{4.5cm}}
\hline
\textbf{Data Flow} & \textbf{PriMod4AI (GPT-OSS)} & \textbf{PriMod4AI (LLaMA3.1)} & \textbf{PILLAR} \\
\hline

DF1  & L, I, NC, DI, U, NR \textbf{[6]} 
     & U, I, DI, NC, L, NR \textbf{[6]}
     & L, I, DI, U, NR \textbf{[5]} \\

DF2  & L, DT, DI, U, NC \textbf{[5]}
     & DI, L, DT, U, NC \textbf{[5]}
     & L, I, DI, U, NR \textbf{[5]} \\

DF3  & L, DT, DI, U, NC \textbf{[5]}
     & U, DT, L, DI, NC \textbf{[5]}
     & L, I, DI, U, NR \textbf{[5]} \\

DF4  & L, I, NR, DT, DI, U, NC \textbf{[7]}
     & L, I, DT, DI, U, NC \textbf{[6]}
     & L, I, DI, U, NR \textbf{[5]} \\

DF5  & L, DT, DI, U, NC \textbf{[5]}
     & U, DT, NC, DI, L \textbf{[5]}
     & L, I, DI, U, NR \textbf{[5]} \\

DF6  & L, I, NR, DT, DI, U, NC \textbf{[7]}
     & I, L, NR, DI, U, DT, NC \textbf{[7]}
     & L, I, NR, U, NC, DI, DT \textbf{[7]} \\

DF7  & L, I, DT, DI, U \textbf{[5]}
     & I, L, DI \textbf{[3]}
     & L, I, DI, U, NR \textbf{[5]} \\

DF8  & L, I, NR, DT, DI, U, NC \textbf{[7]}
     & DI, L, NC, I, DT \textbf{[5]}
     & L, I, NR, U, NC, DI \textbf{[6]} \\

DF9  & L, I, DT, DI, U, NC \textbf{[6]}
     & DI, I, NC, L, DT, U \textbf{[6]}
     & L, I, NR, U, NC, DI \textbf{[6]} \\

DF10 & L, I, DT, DI, U \textbf{[5]}
     & U, DT, L, DI \textbf{[4]}
     & L, I, NR, U, NC, DI \textbf{[6]} \\

DF11 & L, I, DT, DI, U, NC \textbf{[6]}
     & DI, NC, L, DT, I \textbf{[5]}
     & L, I, NR, U, NC, DI, DT \textbf{[7]} \\

DF12 & L, NR, DT, DI, U, NC \textbf{[6]}
     & L, DI, U, NR, DT \textbf{[5]}
     & L, I, NR, U, NC, DI, DT \textbf{[7]} \\

DF13 & L, I, NC, DI, U \textbf{[5]}
     & DI, L, U, I, NC \textbf{[5]}
     & L, I, NR, U, NC, DI \textbf{[6]} \\

DF14 & L, I, NC, DI, DT, U \textbf{[6]}
     & DI, L, U, I, NC \textbf{[5]}
     & L, I, DI, U, NR \textbf{[5]} \\

\hline
\end{tabular}
\end{table*}



\subsection{ Layer B: Combined Threat Space and Model-Centric Analysis}\label{sec:model_centric}
Unlike classical LINDDUN threats, model-centric privacy risks originate from the behaviour of trained models and do not correspond to specific data flows or system components. These risks typically arise from interactions with model parameters, training distributions, latent representations, or generative capabilities, and therefore require an analysis framework that extends beyond the structural boundaries of a DFD. To account for these phenomena, PriMod4AI extracts AI-specific threat descriptions generated by the LLMs and consolidates them into a set of canonical model-centric threat categories through clustering.

Because GPT-OSS and LLaMA-3.1 often describe conceptually similar attacks using different surface forms, for example “AI-Generated Misinformation Using Location Data” and “AI-Fabricated Location Misinformation”, so clustering step is essential for producing a unified taxonomy. This process ensures that semantically equivalent threats across systems (face authentication and autonomous driving) and across model variants are represented consistently, even when the LLMs use divergent wording.

The resulting canonical categories capture well-established families of model-centric privacy risks, such as membership inference attacks, model inversion attacks, reconstruction and leakage mechanisms, embedding or template exposure, dataset replication, and model-generated misleading outputs. In the subsequent analysis, we report how many such canonical categories each PriMod4AI variant identifies in each system and examine their overlap. 

Table~\ref{tab:model_centric_coverage_updated} reports the number of distinct threat expressions and their clustered canonical categories identified for each system. PriMod4AI (GPT-OSS) consistently identifies a wider range of model-centric threats, whereas the LLaMA variant returns more compact but semantically aligned sets. The analysis demonstrates that PriMod4AI extends privacy reasoning beyond traditional design-time taxonomies by uncovering behavioural vulnerabilities specific to machine learning models.

\begin{table}[htbp]
\centering
\scriptsize
\caption{Model-centric threat coverage across systems. 
Canonical Categories represent clusters of semantically related AI-specific privacy threats identified by each model.}
\label{tab:model_centric_coverage_updated}
\begin{tabular}{ p{2.2cm} l  p{1.5cm} }
\toprule
\textbf{System} & \textbf{Model} & \textbf{Canonical Categories} \\
\midrule
\multirow{3}{*}{Autonomous Driving} 
  & PriMod4AI (GPT-OSS) & 11 \\
  & PriMod4AI (LLaMA)   & 7  \\
\midrule
\multirow{3}{*}{Face Authentication}
  & PriMod4AI (GPT-OSS) & 9 \\
  & PriMod4AI (LLaMA)   & 5 \\
\bottomrule
\end{tabular}
\end{table}

\subsection{AI Model-Centric Privacy Attack Knowledge Base (AI\_Privacy\_KB)}\label{sec:AIprivacyKB}

The AI\_Privacy\_KB presented in this appendix provides the complete catalogue of AI-specific, model-centric privacy threats derived from the systematic literature review described in the main paper. The review covered 30 peer-reviewed publications, standards, and authoritative reports published between 2016 and 2025, each offering documented evidence of privacy attacks or vulnerabilities in modern AI systems. While the main text outlines the construction methodology and the integration of this knowledge base within PriMod4AI, this appendix presents the final, structured representation of all identified threats.

Each entry in the AI\_Privacy\_KB corresponds to a single threat extracted from the reviewed sources. To ensure consistency and reproducibility, all threats were encoded using a unified JSON schema that captures:
(i) a unique threat identifier,
(ii) the normalized threat name,
(iii) the associated AI lifecycle stage,
(iv) the relevant flow type used in PriMod4AI’s reasoning,
(v) a concise short description of the attack mechanism,
(vi) a justification of its privacy relevance, and
(vii) the full bibliographic reference of the originating publication.

All threats in this appendix were processed through a unified pipeline including extraction, terminology normalization, deduplication, and harmonization of lifecycle and flow annotations, ensuring coherent and comparable representations across diverse privacy attack types.

A simplified example of the encoding format is shown below Listing\ref{lst:AIPRIVACYKB}:
\begin{lstlisting}[language=json, numbers=none, caption={Example entry from the \texttt{AI\_Privacy\_KB} illustrating the standardized JSON schema used for all model-centric privacy threats.}, label={lst:AIPRIVACYKB}]
{
  "threatId": "6",
  "privacyThreatName": "Inference Attacks on Model Outputs",
  "flowType": "Output flow",
  "aiLifecycleStage": "Inference",
  "shortDescription": "Attackers infer sensitive data from the model's predictions or outputs.",
  "privacyThreatJustification": "Can reveal personal information, even if the data was anonymized or protected during training.",
  "reference": {
    "type": "article",
    "title": "Membership Inference Attacks Against Machine Learning Models",
    "authors": ["R. Shokri",
          "Marco Stronati",
          "Congzheng Song",
          "Vitaly Shmatikov"],
    "journal": "2017 IEEE Symposium on Security and Privacy (SP)",
    "year": "2016",
    "pages": "3-18",
    "url": ""https://api.semanticscholar.org/CorpusID:10488675""
  }
}
\end{lstlisting}
 Tables~\ref{tab:ai_privacy_appendix1} and \ref{tab:ai_privacy_appendix2} present the complete AI\_Privacy\_KB in its final tabular form, consolidating threats from all 30 reviewed publications. This appendix serves as the authoritative source for AI-specific threats used in PriMod4AI and supports transparency, reproducibility, auditing, and future extension of the knowledge base.



\begin{table*}[!t]
\caption{ This table lists AI-specific privacy threats extracted from reviewed and standardized sources, presented in their simplified AI\_Privacy\_KB format.}

\label{tab:ai_privacy_appendix1}
\centering
\footnotesize
\renewcommand{\arraystretch}{1.12}
\begin{tabularx}{\textwidth}{c p{3.2cm} X p{6.2cm}}
\toprule
\textbf{ID} & \textbf{Threat} & \textbf{Short description} & \textbf{Reference} \\
\midrule
1  & Data Quality Compromise & In processing flows during data cleaning and preprocessing, accidental or intentional degradation of data quality leads to unreliable models that mishandle sensitive data. & C.~Sillaber, C.~Sauerwein, A.~Mussmann, and R.~Breu, ``Data Quality Challenges and Future Research Directions in Threat Intelligence Sharing Practice,'' in \textit{Proc. 2016 ACM Workshop on Information Sharing and Collaborative Security (WISCS '16)}, 2016, pp.~65--70. \\
2  & Label Tampering & In processing flows during data labeling, malicious modification of labels causes misclassification that can leak or misrepresent sensitive attributes. & R.~Sharma, G.~K.~Sharma, and M.~Pattanaik, ``Adversarial Label Flipping Attack on Supervised Machine Learning-Based HT Detection Systems,'' in \textit{Proc. 2024 IEEE Int. Symp. Circuits and Systems (ISCAS)}, 2024, pp.~1--5. \\
3  & Inference Attacks on Model Outputs & In output flows during inference, attackers analyze model predictions to infer sensitive data that may have been used for training. & R.~Shokri, M.~Stronati, C.~Song, and V.~Shmatikov, ``Membership Inference Attacks Against Machine Learning Models,'' in \textit{Proc. 2017 IEEE Symp. Security and Privacy (SP)}, 2017, pp.~3--18. \\
4 & Adversarial Attacks & In model-related flows during training and inference, carefully crafted perturbations cause misclassification and can enable privacy and security violations. & X.~Yuan, P.~He, Q.~Zhu, and X.~Li, ``Adversarial Attacks and Defenses in Deep Learning,'' \textit{IEEE Trans. Neural Netw. Learn. Syst.}, vol.~30, no.~9, pp.~2805--2824, 2019. \\
5 & Model Inversion Attacks & In model-related flows during inference, adversaries exploit model outputs to reconstruct sensitive features or entire training records. & M.~Fredrikson, S.~Jha, and T.~Ristenpart, ``Model Inversion Attacks that Exploit Confidence Information and Basic Countermeasures,'' in \textit{Proc. ACM Conf. Computer and Communications Security (CCS)}, 2015. \\
6 & Data Poisoning Attacks & In data collection and model-related flows during data collection and training, adversaries tamper with data to introduce backdoors or targeted errors. & E.~Bagdasaryan, A.~Veit, Y.~Hua, D.~Estrin, and V.~Shmatikov, ``How To Backdoor Federated Learning,'' in \textit{Proc. Int. Conf. Artificial Intelligence and Statistics (AISTATS)}, 2020, pp.~2938--2948. \\
7 & Membership Inference Attacks & In model-related and output flows during training and inference, attackers test whether specific records were part of the training dataset. & R.~Shokri, M.~Stronati, C.~Song, and V.~Shmatikov, ``Membership Inference Attacks Against Machine Learning Models,'' in \textit{Proc. 2017 IEEE Symp. Security and Privacy (SP)}, 2017, pp.~3--18. \\
8 & Risk of Non-Compliance & Across data collection, processing, and deployment flows, AI systems fail to meet privacy regulations such as the GDPR, leading to legal and trust risks. & S.~Shahriar, S.~Allana, S.~M.~Hazratifard, and R.~Dara, ``A Survey of Privacy Risks and Mitigation Strategies in the Artificial Intelligence Life Cycle,'' \textit{IEEE Access}, vol.~11, pp.~61829--61854, 2023. \\
9 & AI-Assisted Hacking & In processing flows during inference, AI tools generate or support cyberattacks, lowering the barrier to sophisticated intrusions. & ``From ChatGPT to ThreatGPT: Impact of Generative AI in Cybersecurity and Privacy,'' \textit{IEEE Access}, 2023. \\
10 & Deep Leakage from Gradients & In model-related flows during centralized or federated training, shared gradients leak enough information to reconstruct training examples. & L.~Zhu, Z.~Liu, and S.~Han, ``Deep Leakage from Gradients,'' in \textit{Proc. 33rd Int. Conf. Neural Information Processing Systems (NeurIPS)}, 2019. \\
11 & Model Extraction via Prediction APIs & In output flows during deployment and inference, black-box querying of prediction APIs is used to clone models and their behavior. & F.~Tramèr, F.~Zhang, A.~Juels, M.~K.~Reiter, and T.~Ristenpart, ``Stealing Machine Learning Models via Prediction APIs,'' in \textit{Proc. USENIX Security Symp.}, 2016. \\
12 & Training Data Extraction from LLMs & In output flows during inference and deployment, adversaries craft prompts that cause LLMs to reveal training data.  & N.~Carlini \textit{et al.}, ``Extracting Training Data from Large Language Models,'' in \textit{Proc. 30th USENIX Security Symp.}, 2021. \\
13  & Side-Channel Attacks & In model-related flows during deployment, adversaries exploit timing, power, or other side channels to infer secrets without direct access to model data. & B.~I.~Priya, P.~V.~R.~D.~P.~Rao, and D.~V.~L.~Parameswari, ``Shielding secrets: developing an enigmatic defense system with deep learning against side channel attacks,'' \textit{Discov. Sustain.}, vol.~5, art.~249, 2024, doi: 10.1007/s43621-024-00455-4. \\
14 & Reconstruction Attacks & In model-related and output flows during data processing and deployment, attackers reconstruct private datasets using model outputs and auxiliary information. & S.~Shahriar, S.~Allana, S.~M.~Hazratifard, and R.~Dara, ``A Survey of Privacy Risks and Mitigation Strategies in the Artificial Intelligence Life Cycle,'' \textit{IEEE Access}, vol.~11, pp.~61829--61854, 2023, doi: 10.1109/ACCESS.2023.3287195. \\
15 & Exfiltration via Cyber Means & In data collection, training, and deployment flows, attackers use network or physical compromise to steal datasets and model weights & ``A Survey on Privacy Attacks Against Digital Twin Systems in AI-Robotics,'' \textit{arXiv}, 2024. [Online]. Available: \url{https://arxiv.org/abs/2406.18812} \\
\bottomrule
\end{tabularx}
\end{table*}

\begin{table*}[!t]
\caption{ This table lists AI-specific privacy threats extracted from reviewed and standardized sources, presented in their simplified AI\_Privacy\_KB format.}
\label{tab:ai_privacy_appendix2}
\centering
\footnotesize
\renewcommand{\arraystretch}{1.12}
\begin{tabularx}{\textwidth}{c p{3.2cm} X p{6.2cm}}
\toprule
\textbf{ID} & \textbf{Threat} & \textbf{Short description} & \textbf{Reference} \\
\midrule
16  & AI-Generated Misinformation & During output flows at deployment and inference, AI systems generate false or misleading content that can harm reputations and privacy. & H.-P.~Lee, Y.-J.~Yang, T.~S.~von Davier, J.~Forlizzi, and S.~Das, ``Deepfakes, Phrenology, Surveillance, and More! A Taxonomy of AI Privacy Risks,'' \textit{arXiv preprint arXiv:2310.07879}, 2023. [Online]. Available: \url{https://arxiv.org/abs/2310.07879} \\
17  & AI-Enabled Social Engineering & During output flows at deployment and inference, AI generates highly personalized phishing or manipulation messages that increase the risk of disclosing sensitive data. & H.-P.~Lee \textit{et al.}, same as ID 1 above. \\
18  & Critical Data Removal & In processing flows during data cleaning and preprocessing, selective removal of critical data points distorts model behavior and may indirectly expose personal information. & V.~Bazarevsky \textit{et al.}, ``BlazeFace: Sub-millisecond Neural Face Detection on Mobile GPUs,'' \textit{arXiv preprint arXiv:1907.05047}, 2019. \\
19  & Biased Data Processing & In processing flows during data preprocessing, biased transformations amplify unfairness and can result in discriminatory handling of personal data. & L.~E.~Celis, V.~Keswani, and N.~K.~Vishnoi, ``Data preprocessing to mitigate bias: A maximum entropy based approach,'' \textit{arXiv preprint arXiv:1906.02164}, 2020. \\
20  & Data Leakage During Preprocessing & In processing and model-related flows across preprocessing, training, and testing, poor handling of notebooks and pipelines causes unintended exposure of sensitive information. & C.~Yang \textit{et al.}, ``Data Leakage in Notebooks: Static Detection and Better Processes,'' \textit{arXiv preprint arXiv:2209.03345}, 2022. \\
21 & Privacy Leakage During Monitoring & In output flows during monitoring, logs, metrics, or traces recorded for performance tracking inadvertently expose sensitive data. & M.~Jegorova \textit{et al.}, ``Survey: Leakage and Privacy at Inference Time,'' \textit{arXiv preprint arXiv:2107.01614}, 2022. \\
22 & Poisoning Attacks & In model-related flows during training, attackers inject malicious samples into datasets to corrupt models and enable privacy-relevant misbehavior. & C.~Sitawarin \textit{et al.}, ``A Survey on Data Poisoning Attacks in Machine Learning,'' \textit{arXiv preprint arXiv:2301.05412}, 2023. \\
23 & Data Extraction Attacks & In model-related and output flows during inference, adversaries design queries to extract sensitive training data from model outputs. & N.~Carlini \textit{et al.}, ``Extracting Training Data from Large Language Models,'' \textit{arXiv preprint arXiv:2012.07805}, 2021. \\
24 & Ethical and Societal Risks & In model-related flows during deployment and inference, AI systems may impact societal norms, autonomy, and rights, creating systemic privacy harms. & ``Privacy Risks of General Purpose AI Systems: A Foundation for Investigation Practitioner Perspectives,'' \textit{arXiv preprint arXiv:2407.02027}, 2024. \\
25 & LLM Data Leakage & In output flows during inference, large language models inadvertently reveal private, proprietary, or training data in generated responses. & ``A Survey on Privacy Attacks Against Digital Twin Systems in AI-Robotics,'' \textit{arXiv preprint arXiv:2406.18812}, 2024. \\
26 & Synthetic Data Inference & In data collection and model-related flows during processing and inference, attackers use synthetic data statistics to re-identify individuals from the original dataset. & C.~Zhang, ``State-of-the-Art Approaches to Enhancing Privacy Preservation of Machine Learning Datasets: A Survey,'' \textit{arXiv preprint arXiv:2404.16847}, 2025. \\
27 & Unauthorized AI Tool Usage & Across all flows and stages, employees use unapproved AI tools, bypassing security controls and data governance. & Forbes Technology Council, ``Emerging AI Threats To Navigate In 2025 And Beyond,'' \textit{Forbes}, Feb. 2025. [Online]. Available: \url{https://www.forbes.com/councils/forbestechcouncil/2025/02/12/emerging-ai-threats-to-navigate-in-2025-and-beyond/} \\
28 & API-Based Model Stealing & In output flows during deployment, repeated queries to model APIs are used to reconstruct or approximate proprietary models containing privacy-sensitive patterns. & SecureSustain, ``International AI Safety Report 2025 -- Security \& Sustainability,'' 2025. [Online]. Available: \url{https://securesustain.org/report/international-ai-safety-report-2025/} \\
29 & Re-identification of Anonymized Data & In processing and inference stages, AI techniques re-link anonymized data with external sources, re-identifying individuals. & ``Artificial Intelligence and Privacy: Examining the Risks and Potential Solutions,'' \textit{Artificial Intelligence}, 2024. [Online]. Available: \url{https://www.researchgate.net/publication/378545816} \\
30 & Malicious Code Generation & In output flows during inference, AI systems generate malware, ransomware, or exploit scripts that can be used to compromise privacy. & ``From ChatGPT to ThreatGPT: Impact of Generative AI in Cybersecurity and Privacy,'' \textit{IEEE Access}, 2023, doi: 10.1109/ACCESS.2023.3300381. \\
\bottomrule
\end{tabularx}
\end{table*}

\end{document}